\documentclass{aa}

\usepackage{natbib}
\usepackage{subfig}
\bibpunct{(}{)}{;}{a}{}{,} 
\usepackage{upgreek}
\usepackage{graphicx}
\usepackage{txfonts}
\usepackage[urlcolor=blue,colorlinks=true,  allcolors=blue ]{hyperref}
\usepackage{threeparttablex} 
\usepackage{lscape} 
\usepackage{url}
\usepackage{tablefootnote}
\usepackage{multicol}

\usepackage{xcolor}
\usepackage[normalem]{ulem}

\def\deg  {\ifmmode {^\circ}\else {$^\circ$}\fi}

\definecolor{malachite}{rgb}{0.34, 0.7, 0.22}

\def\cof {$^{12}\mathrm{CO}~(J=1\rightarrow0)$}

\def\msun{$M_{\odot}$}
\def\ag{$A _\mathrm{G}$}

\def\kms     {km~s$^{-1}$}
\newcommand{\HII}{\mbox{H\,\textsc{ii}}}%

\def\msun{$M_{\odot}$}
 
\def\ag{$A_{G}$}
\def\av{$A_{V}$}
\def\cofs {$^{12}\mathrm{CO}$}

\def\deg  {\ifmmode {^\circ}\else {$^\circ$}\fi}

\begin{document}
 
    \title{Distances to molecular clouds in the second Galactic quadrant}
 
   \author{Qing-Zeng Yan, Ji Yang, Yan Sun, Yang Su, Ye Xu, Hongchi Wang, Xin Zhou, Chen Wang}
   \institute{Purple Mountain Observatory and Key Laboratory of Radio Astronomy,  Chinese Academy of Sciences, 10 Yuanhua Road, Qixia District, Nanjing 210033, People's Republic of China. \email{jiyang@pmo.ac.cn,qzyan@pmo.ac.cn}
 }

   \date{}
   
\titlerunning{Molecular cloud distances}
\authorrunning{Yan et al}

  \abstract
   {We present distances to 76 medium-sized molecular clouds and an extra large-scale one in the second Galactic quadrant ($104\fdg75  <l<150\fdg25 $ and $|b|<5\fdg25$), 73 of which are accurately measured for the first time. Molecular cloud samples are drawn from $l$-$b$-$V$ space ($-95 < V_{\rm LSR}< 25$ \kms) with the density-based spatial clustering of applications with noise (DBSCAN) algorithm, and distances are measured with the background-eliminated extinction-parallax (BEEP) method using extinctions and Gaia DR2 parallaxes. The range of measured distances to 76 molecular clouds is from 211 to 2631 pc, and the extra large-scale molecular cloud appears to be a coherent structure at about 1 kpc, across about 40\deg\ ($\sim$700 pc) in the Galactic longitude. } 

   \maketitle
%


 

\section{Introduction} \label{sec:intro} 


As a basic parameter, the distance is pivotal to the investigation of physical properties of molecular clouds \citep{2012ARA&A..50..531K,2015ARA&A..53..583H}, as well as their internal structures \citep{2018A&A...619A.106G,2020A&A...638A..85R} and the distribution in the Milky Way \citep{2019ApJ...885..131R,2020Natur.578..237A}.  The accuracy of  distances to high-mass star-forming regions has been improved significantly due to the progress of the very-long-baseline interferometry (VLBI) technique \citep{2006Sci...311...54X}. For instance, the BeSSeL Survey \footnote{\href {http://bessel.vlbi-astrometry.org/}{http://bessel.vlbi-astrometry.org/}} \citep{2014ApJ...783..130R,2019AJ....157..200Z,2019ApJ...874...94W} and  the VERA (VLBI Exploration of Radio Astrometry) project  \footnote{\href {https://www.miz.nao.ac.jp/veraserver/}{https://www.miz.nao.ac.jp/veraserver/}} \citep{2003ASPC..306..367K,2007PASJ...59..897H, 2012PASJ...64..136H, 2020PASJ...72...50V} have collectively measured  parallaxes of approximately 200 molecular masers toward high-mass star-forming regions with high precision \citep{2019ApJ...885..131R}.  With the release of a large amount of stellar parallax and extinction measurements in the Gaia DR2 catalog \citep{2016A&A...595A...1G,2018A&A...616A...1G}, distances to many local maser-free molecular clouds can also be determined accurately.

At high Galactic latitudes, the dust environment is relatively simple,  and measuring distances to molecular clouds  with stellar extinction is straightforward. For instance, \citet{2019ApJ...879..125Z} and \citet{2020A&A...633A..51Z} determined distances to many molecular clouds and star-forming regions at high Galactic latitudes ($|b|>10\deg$), and \citet{2019A&A...624A...6Y} derived distances to $\sim$50 molecular clouds at high Galactic latitudes as well. The systematic error in those distances is about 5\%. In addition to accurate distances, large solid angles of molecular clouds at high Galactic latitudes allow us to investigate their three-dimensional (3-D) structures and motions. For instance, \citet{2018A&A...619A.106G} studied the 3-D shape of Orion A and found that it resembles large-scale filaments in the Milky Way. \citet{2020A&A...638A..85R} studied the Taurus star-forming region with Gaia DR2 parallaxes and proper motions and found that distances to its stellar members are in the range of [130, 160] pc, and they speculated that Taurus B is moving toward Taurus A. \citet{2020Natur.578..237A} studied the structure of local molecular clouds, and found that those molecular clouds show a sinusoidal wave (``Radcliffe Wave'') instead of a ring in 3-D space.


However, due to the complexity of the interstellar medium, distances to maser-free molecular clouds in the Galactic plane are relatively difficult to measure. In the Galactic plane, the present of multiple components along the line of sight and the velocity crowding in the inner Galaxy makes  molecular cloud identification difficult. Consequently, matching dust cloud distances \citep{2019ApJ...887...93G,2020MNRAS.493..351C,2020MNRAS.496.4637C} to molecular clouds is not straightforward and high-quality spectral images of molecular clouds are needed. In order to deal with molecular cloud distances in the Galactic plane, \citet{2019ApJ...885...19Y} proposed a background-eliminated extinction-parallax (BEEP) method based on the Gaia DR2 data and the  Milky Way Imaging Scroll Painting (MWISP) CO survey. This method reveals the extinction jump caused by molecular cloud using  the extinction difference between on- and off-cloud stars. Using this method, \citet{2019ApJ...885...19Y} derived distances to 11 molecular clouds in the third Galactic quadrant ($209\fdg75  <l< 219\fdg75$ and $|b|<5\deg$), and \citet{2020ApJ...898...80Y} derived distances to 28 molecular clouds in the first Galactic quadrant ($25\fdg8  <l<49\fdg7 $ and $|b|<5\deg$).



In this work, we examine distances to molecular clouds in the second Galactic quadrant ($104\fdg75  <l<150\fdg25 $ and $|b|<5\fdg25$), mapped by the MWISP survey with  high sensitivities and moderate angular resolutions \citep{2016ApJS..224....7D,2017ApJS..229...24D,2020ApJS..246....7S}, including  images of three CO isotopologue lines. For the purpose of distance determination, we only use the \cof\ images to identify molecular clouds with the density-based spatial clustering of applications with noise (DBSCAN) clustering algorithm \citep{2020ApJ...898...80Y}, whose distances are subsequently examined with the  BEEP method \citep{2019ApJ...885...19Y}.

This paper is organized as follows. The next section (Section \ref{sec:data})  describes the CO data, the cloud identification algorithm, and the BEEP method. Section \ref{sec:result} presents distances to  molecular clouds. Discussions about the molecular cloud distance are presented in Section \ref{sec:discuss}, and we summarize the conclusions in Section \ref{sec:summary}.





\section{Data And Methods} 
\label{sec:data}

\subsection{CO data} 

The \cofs\ image of the examined region ($104\fdg75  <l<150\fdg25 $, $|b|<5\fdg25$, and $-95 < V_{\rm LSR}< 25$ \kms) is part of the MWISP\footnote{\href {http://www.radioast.nsdc.cn/mwisp.php}{http://www.radioast.nsdc.cn/mwisp.php}} CO survey \citep{2019ApJS..240....9S}. The MWISP survey maps the Galactic plane ($|b|<5.25$) in the Northern sky tile by tile ($30\arcmin\times30\arcmin$), and this region in the second Galactic quadrant is almost complete except for a few tiles at relatively high Galactic latitudes ( $129\deg  <l<130\deg$ and $b= 4\fdg5$). The rms noise  of  \cofs\ is about 0.5 K, and the angular and velocity resolutions are about 49\arcsec\ and 0.167 \kms, respectively.

\subsection{Molecular cloud identification}
 \label{sec:cloudIdentification}

Before measuring distances, we drew molecular cloud samples using the method  developed by \citet{2020ApJ...898...80Y} based on the DBSCAN clustering algorithm\footnote{\href {https://scikit-learn.org/stable/auto\_examples/cluster/plot\_dbscan.html}{https://scikit-learn.org/stable/auto\_examples/cluster/plot\_dbscan.html}}.

DBSCAN identifies connected structures in position-position-velocity (PPV) space, paying no attention to the internal structure, the morphology, and the scale of molecular clouds.  Without distance information,  we initially assume that close components in PPV space are also close physically, but this assumption can easily be  violated. The molecular cloud boundary depends on the sensitivity of observations, and DBSCAN is insensitive to the morphology of molecular clouds. Consequently, the DBSCAN molecular cloud samples contains multiple kinds of shapes, such as cometary structures, Gaussian sources, and filaments. In other words, DBSCAN only searches separated structures in PPV space.

DBSCAN cloud samples are particularly suitable for distance measurements, and splitting large structures in PPV space is meaningless for the BEEP method, because there is no nearby off-cloud regions for interior regions of molecular clouds.

DBSCAN has two parameters, MinPts and $\epsilon$. $\epsilon$ defines the neighborhood of voxels, while MinPts defines core points, which are voxels whose numbers of neighboring voxels  (including itself) are at least MinPts. In PPV space,  $\epsilon$ corresponds to three connectivity types ($\epsilon=1,\sqrt2,\rm and\ \sqrt3$),  and for large $\epsilon$, MinPts also needs to be high to avoid including too many noises. \citet{2020ApJ...898...80Y} found that for appropriate lower MinPts values, the three connectivity types yield similar cloud samples, so we use connectivity 1 ($\epsilon=1$) and MinPts 4. The minimum cutoff on the PPV data cubes is 2$\sigma$ ($\sim$1 K).

Due to the low cutoff (2$\sigma$), noises are inevitable with connectivity 1 and MinPts 4, and post selection criteria are applied to remove noise DBSCAB clusters. Here, we follow \citet{2020ApJ...898...80Y} using four criteria, two of which are related to sensitivity: (1) the minimum voxel number is 16 and (2) the minimum peak brightness temperature is 5$\sigma$. The rest two criteria  are related to resolution: (3) the projection area contains a beam (a compact 2$\times$2 region) and (4) the minimum channel number in the velocity axis is 3.


Around $-$24 \kms, the data cube contains a bad channel, yielding sample contamination. Those fake molecular clouds were removed  according to their $l$-$b$-$v$ ranges.


 \begin{figure*}[ht!]
 \centering
 \includegraphics[width=0.95\textwidth]{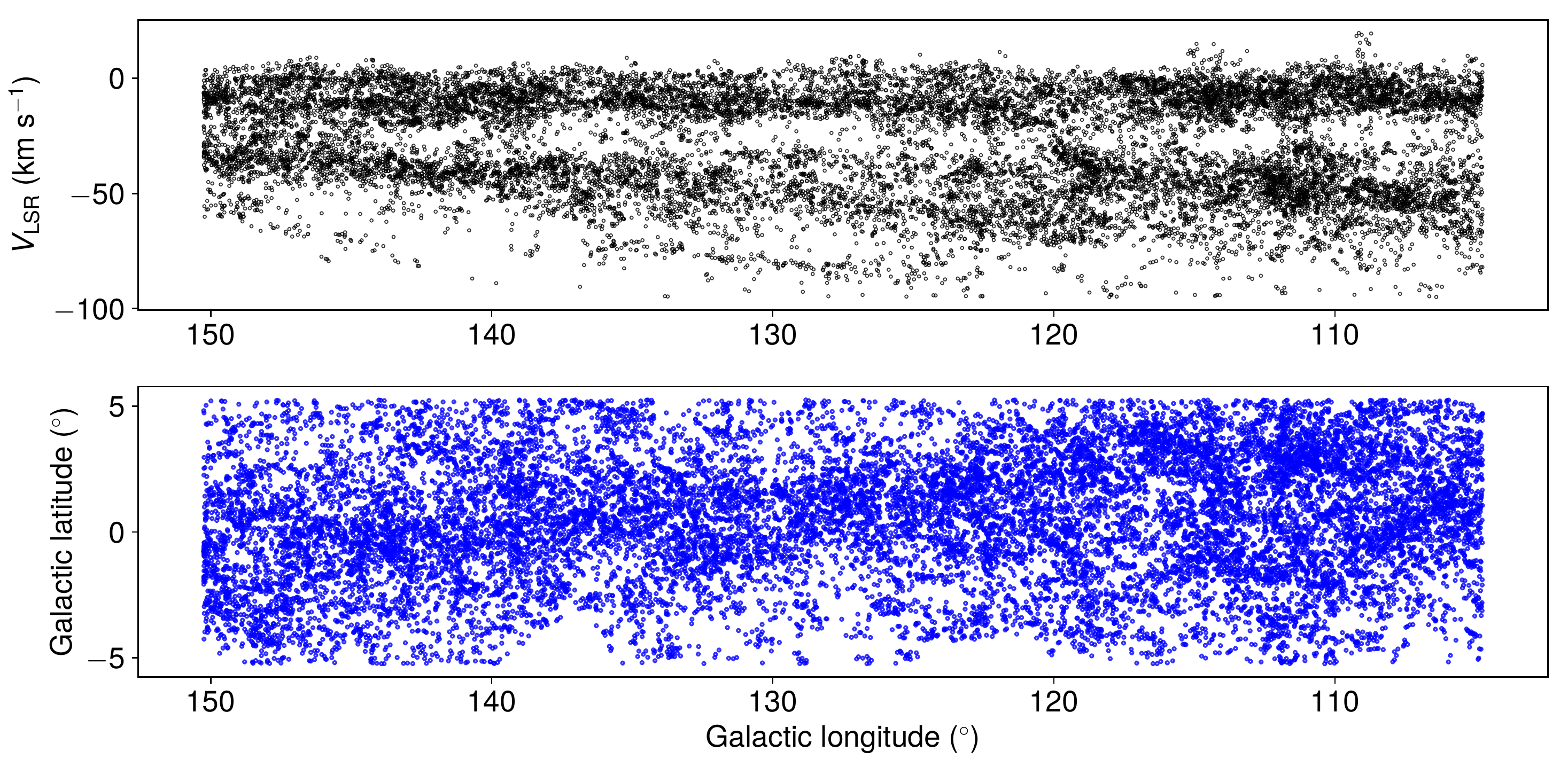} 
 \caption{$l$-$V$ and $l$-$b$ distributions of the 18,413 molecular clouds identified in the second Galactic quadrant. \label{fig:lbv} }
 \end{figure*}

  \begin{figure*}[ht]
  \centering
    \subfloat[]{\includegraphics[width=0.32\textwidth]{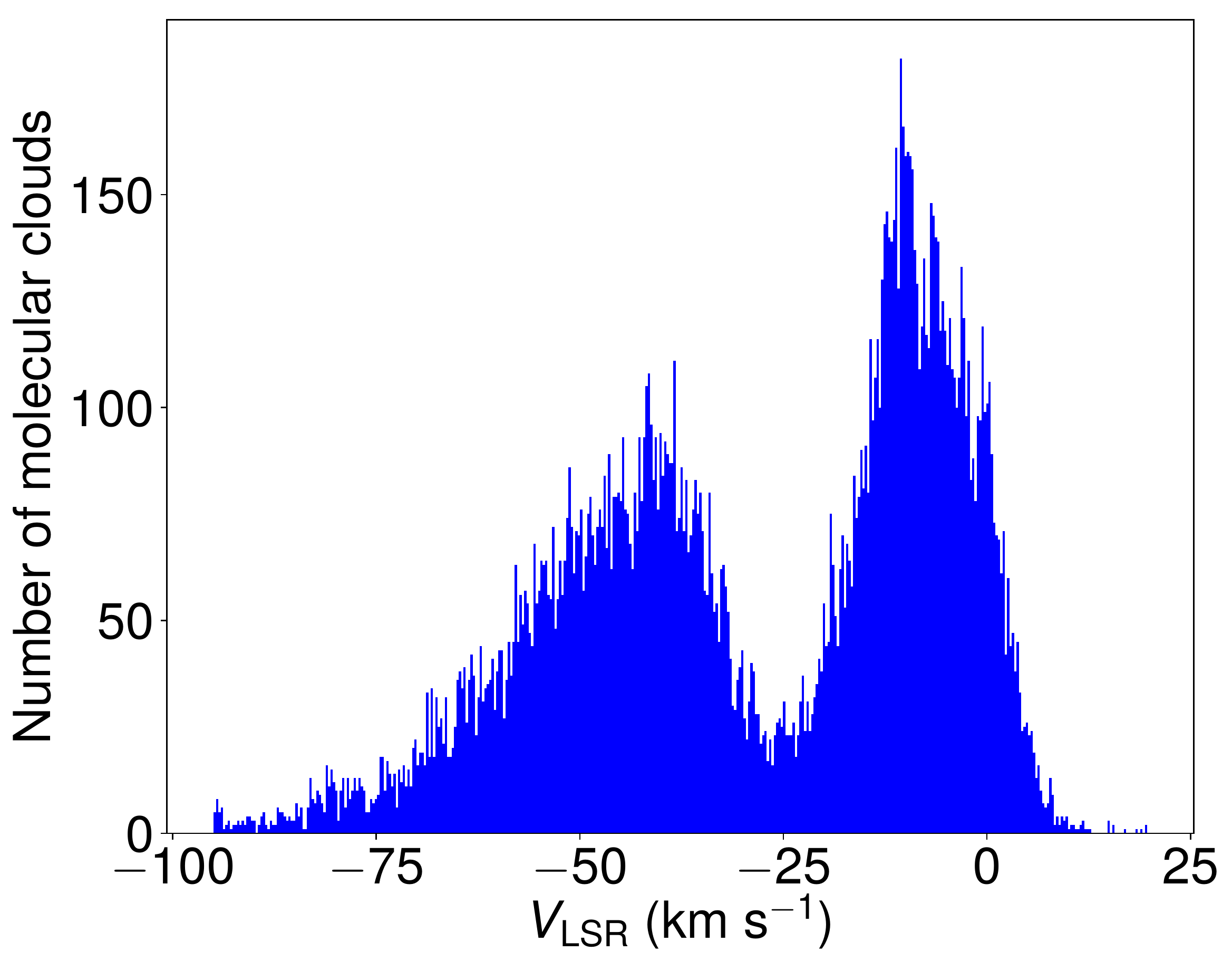}}
  \subfloat[]{\includegraphics[width=0.32\textwidth]{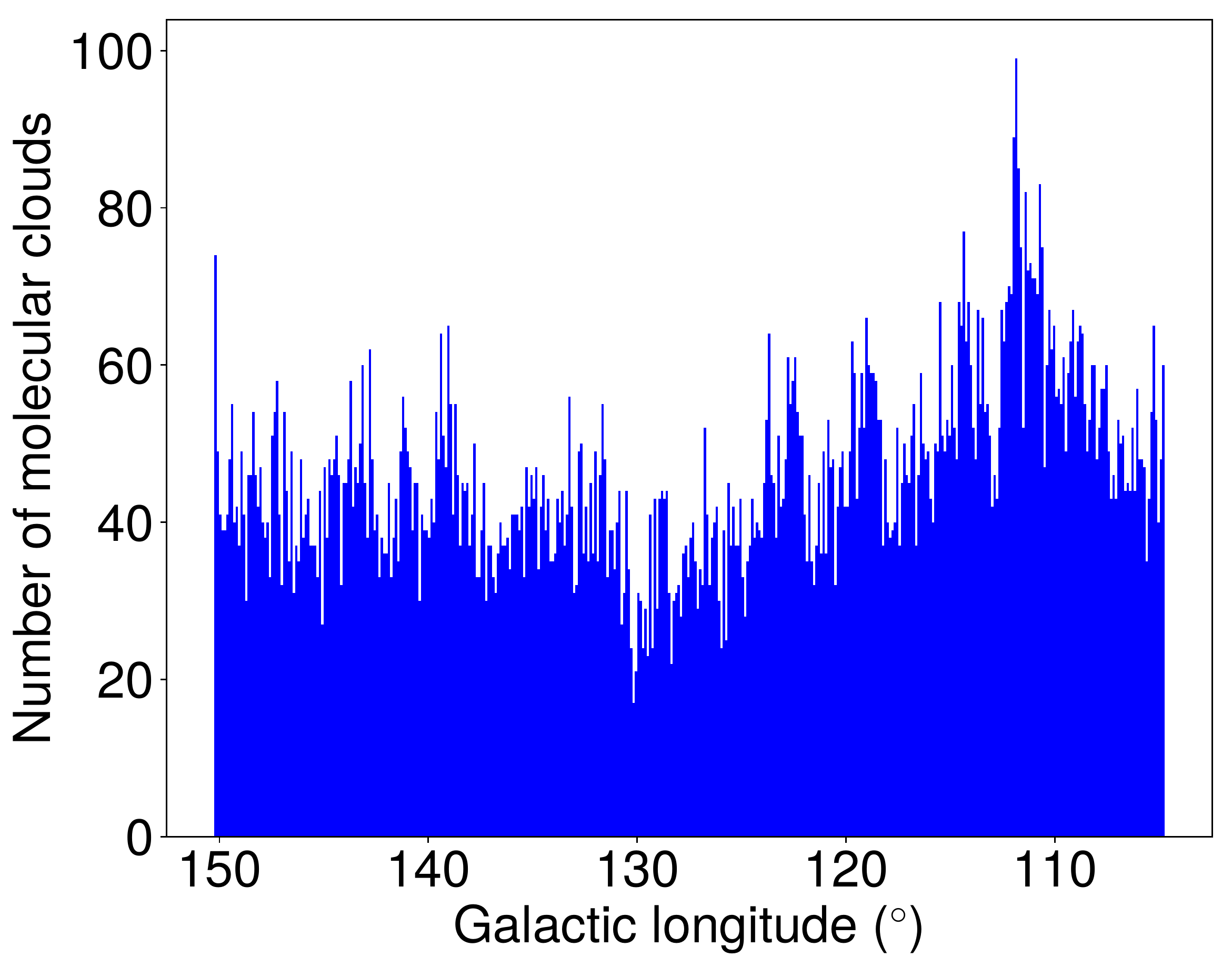}}
  \subfloat[]{\includegraphics[width=0.32\textwidth]{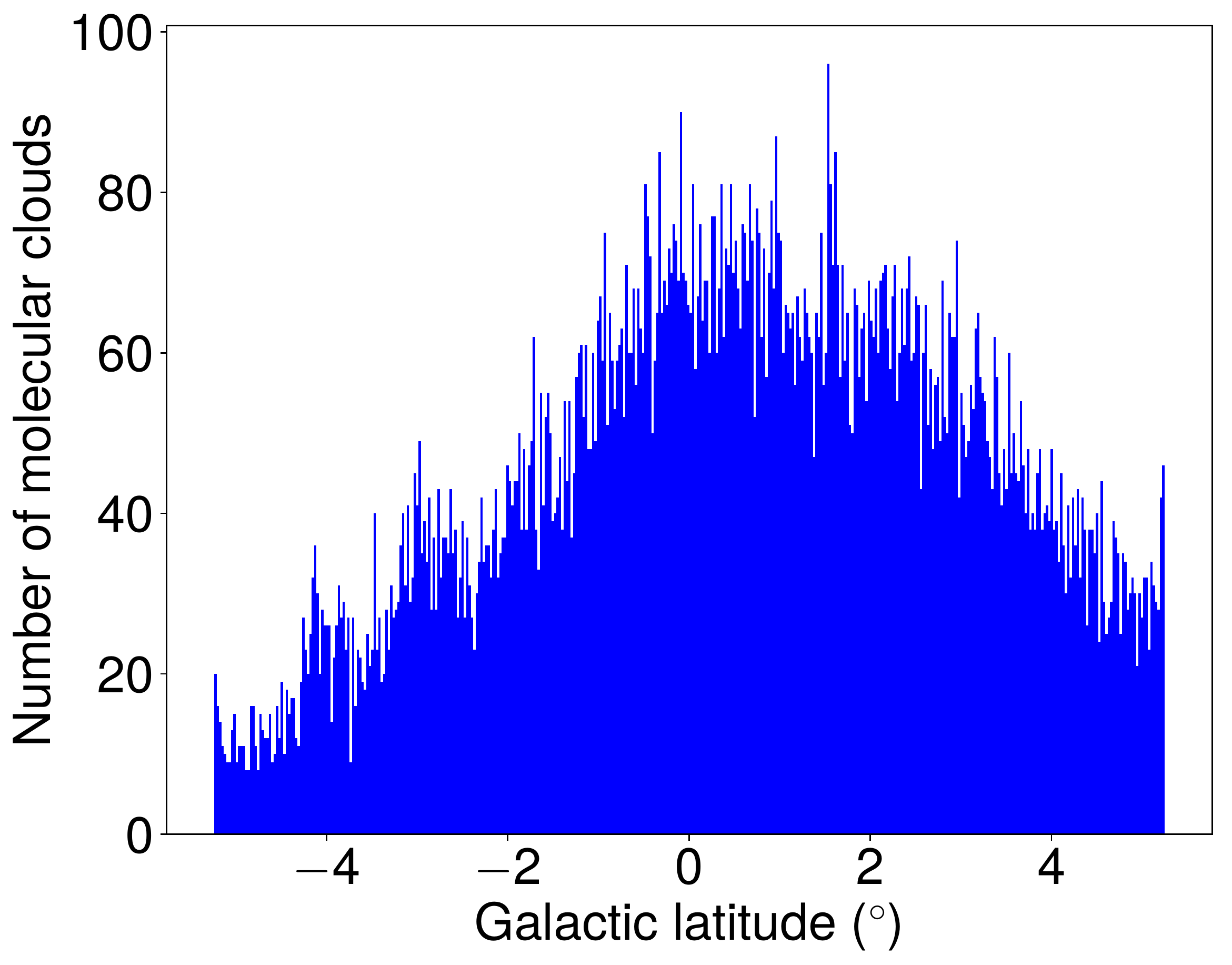}}
  \caption{ Histograms of the 18,413  molecular clouds in the (a) $v$, (b) $l$, and (c) $b$ axes. \label{fig:lbvhist} }
  \end{figure*}








\subsection{Distance measurement}

The procedure of measuring distances using the BEEP method is identical to that of \citet{2020ApJ...898...80Y}. The BEEP method is developed by \citet{2019ApJ...885...19Y} and is used to measure distances to molecular clouds in the Galactic plane, where the dust environment is complex. Along the line of sight, stellar extinction picks up variations that are unrelated to the target molecular clouds, hindering accurate distance measurements. To overcome this difficulty, \citet{2019ApJ...885...19Y} proposed to use off-cloud stars to trace unrelated extinction variation, which should be subtracted from on-cloud stellar extinction.  

 The slow irregular increase of off-cloud extinction with respect to the distance is modeled with the isotonic regression \citep{2019ApJ...885...19Y}. Isotonic regression models data with monotonic functions, and for the stellar extinction, monotonically increasing functions are used. Essentially, the isotonic regression fits the extinction with multiple step functions to grab the overall variation, and the fitted isotonic regression lines serve as extinction baselines, which are subtracted from the extinction of on-cloud stars. For each molecular cloud, all off-cloud stars collectively yield one extinction baseline, and all on-cloud stars commonly use this baseline.

The BEEP method requires the off-cloud region close to the on-cloud region, and both regions need to contain sufficient stars to finely reveal the extinction jump. If the off-cloud region is too far, the dust environment  of off-cloud stars is dissimilar from that of on-cloud stars, and cannot be used for calibration. For large molecular clouds, we manually chose sub-regions that contain both on- and off-cloud regions, i.e. not all on-cloud stars are used. Molecular cloud edges in the sub-regions should be relatively sharp in the integrated intensity map. In addition to the closeness, due to the large uncertainty in the extinction, the numbers of on- and off-cloud stars are both required to be sufficient to trace the  extinction variation reliably. For near ($<$300 pc)  or small molecular clouds, on-cloud stars are usually too few to reveal distances.

A box region needs to contain  both on- and off-cloud stars to measure the molecular cloud distance. The box region is manually chosen, and there are a few useful rules to follow. For a small molecular cloud, a rectangular region that covers the whole molecular cloud and  incorporates a comparable noise region around is adequate. For a large molecular cloud, however, multiple medium-sized sub-regions should be examined, and those with sharply defined boundaries or at relatively high Galactic latitudes are preferable. If a molecular cloud is  both large and near ($<$300 pc), the box region should contain the whole molecular cloud  to use as many on-cloud stars as possible. See Section \ref{sec:result} for three examples of determining box regions in the distance measurements, regarding a large-scale molecular cloud and two molecular clouds with relatively smaller angular areas.

The distance examination procedure starts from the largest molecular clouds. For each molecular cloud, we obtain an integration map in the velocity range [$V_{\rm center}-3\Delta V$, $V_{\rm center}+3\Delta V$], where $V_{\rm center}$ is the average radial velocity weighted by the voxel brightness temperature (the first moment) and $\Delta V$ is the weighted standard deviation (the second moment). On-cloud stars need to be located within the cloud edge defined by DBSCAN, and on the contrary, off-cloud stars are outside of the edge. We examined all molecular clouds (1677 in total) with angular areas  $>$0.015 deg$^2$, which is estimated to be the minimum angular area required for distance measurement \citep{2020ApJ...898...80Y}.


\begin{figure*}[ht!]
   \centering
 \subfloat[]{\includegraphics[width=0.88\textwidth]{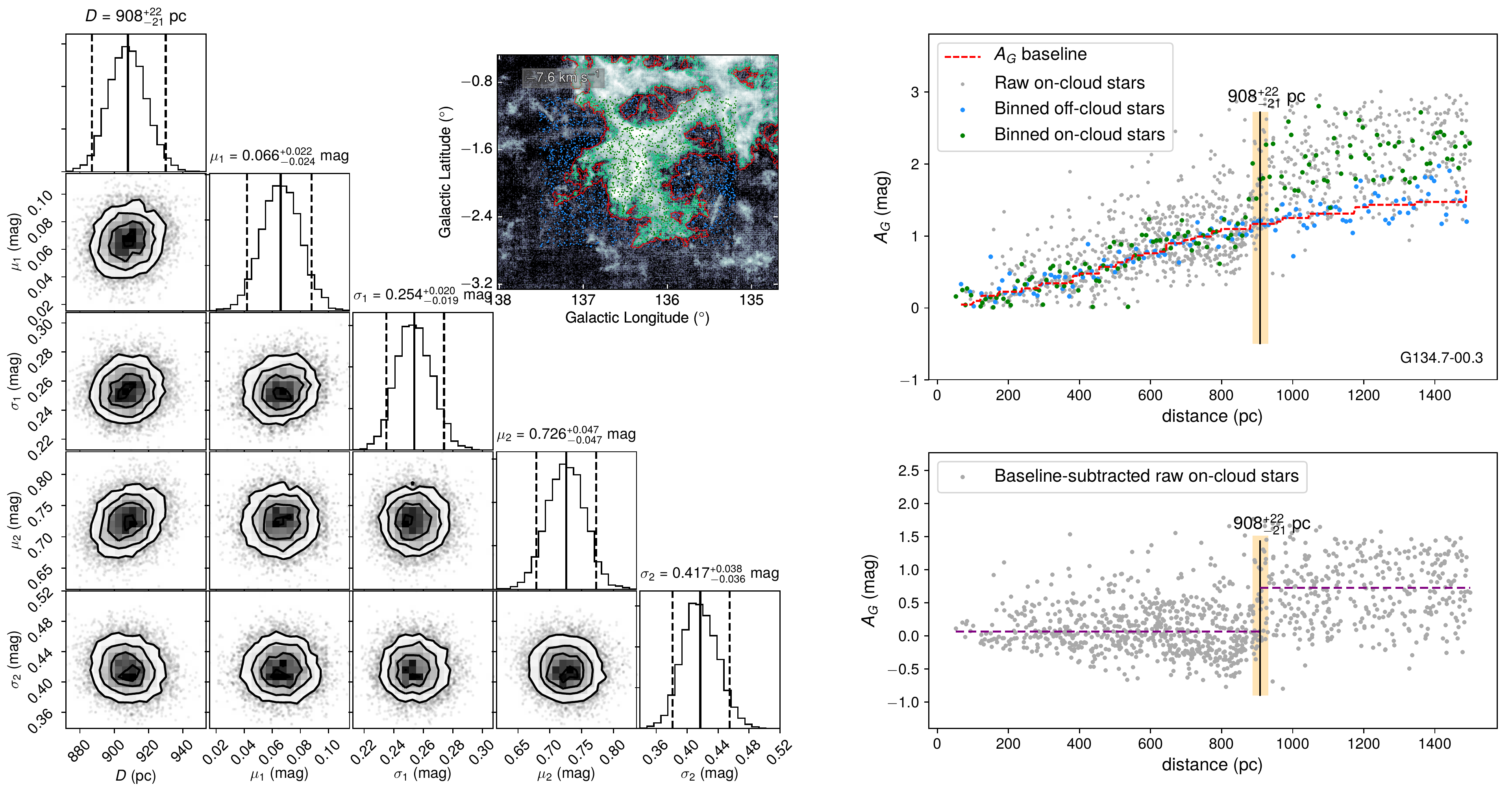}} \\
  \subfloat[]{\includegraphics[width=0.88\textwidth]{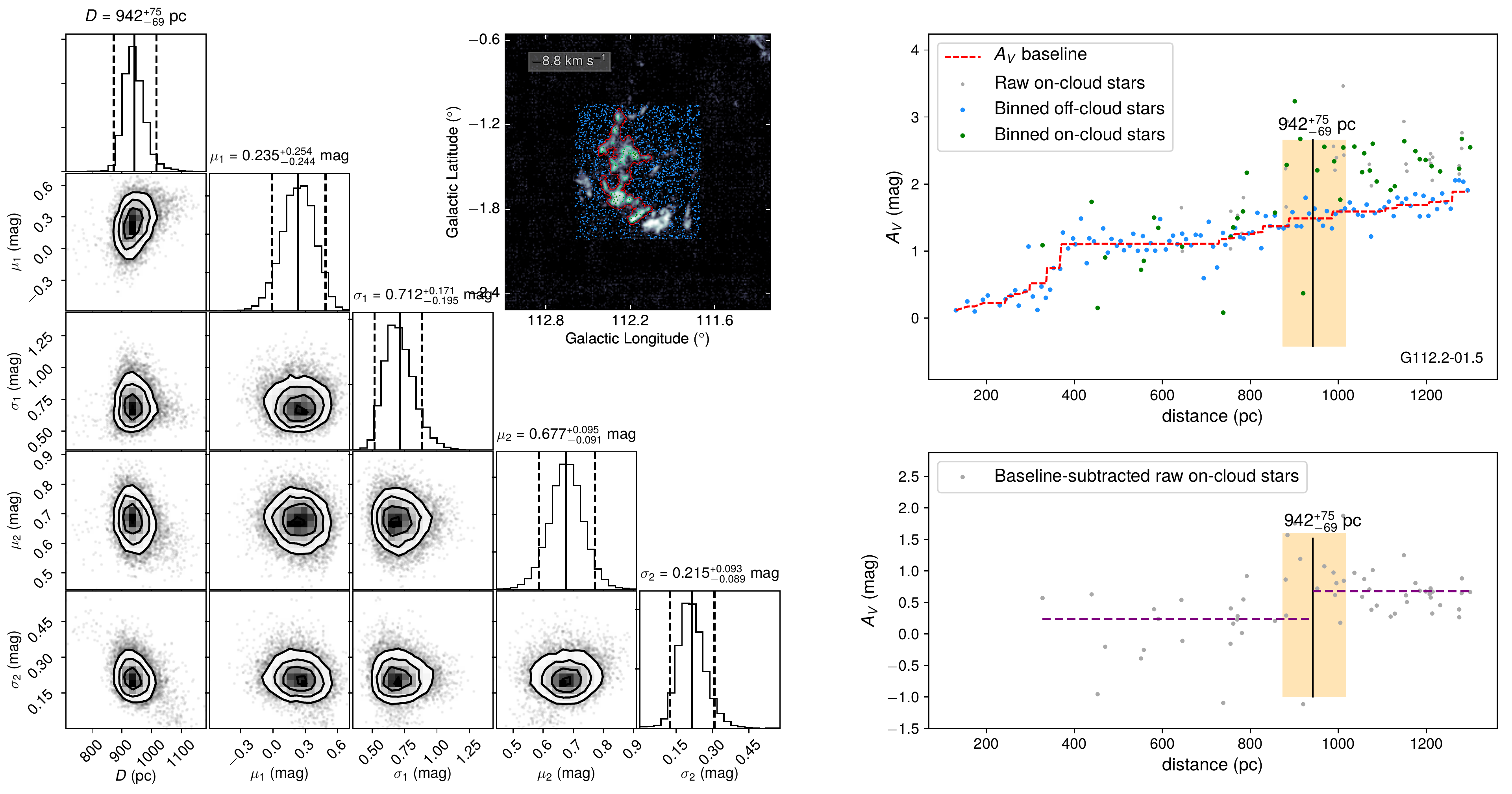}}
\caption{Distances to (a) G134.7$-$00.3 with \ag\  and (b) G112.2$-$01.5 with \av. On the right-hand side plots of both panels, the red-dashed line is the baseline obtained with off-cloud stars, and the broken horizontal purple line is the extinction variation of on-cloud stars after subtracting the baseline. In the CO image insets, the green and light blue points represent samples of on- and off-cloud stars, respectively. The red line delineates the boundary of molecular clouds, while the green line is the lower cutoff of CO emission towards on-cloud stars. In the CO image insets of panel (a), the off-cloud region is selected to be close to the on-cloud region to accurately remove irregular extinction variations of on-cloud stars. The left corner maps are the MCMC sampling results of five parameters: the distance ($D$), the mean and standard deviation of foreground extinction ($\mu_1$ and $\sigma_1$), and the mean and standard deviation of background extinction ($\mu_2$ and $\sigma_2$). The error of distances is the 95\% highest posterior densities (HPD), about two standard deviations for Gaussian distributions. See \citet{2019ApJ...885...19Y} for other details.  \label{fig:twoclouds} }
\end{figure*}

The CO emission toward on-cloud stars (signal level) need to be significant to reveal the extinction jump. The signal level used is 4 K \kms, and the manual choice of signal level is one source of the systematic  distance error. The signal level changes the average extinction of background on-cloud stars, thereby changing the distance results, but for molecular clouds that show clear extinction jumps, the effect of signal level is insignificant.  The CO emission toward off-cloud stars (noise level) need to be low, and the upper limit is 1 K \kms.

We only keep those molecular clouds that show unique and evident extinction jumps. When \ag\ stars are insufficient, we try \av\ measurements \citep{2019A&A...628A..94A}. In order to yield reliable distances, all flags of \av\ data are required to be 0. To make the results consistent, we calibrated  systematic errors in Gaia DR2 following the procedure of \citet{2019A&A...628A..94A}. The distance is modeled as a switch point of two Gaussian distributions, from the foreground stellar extinction distribution (a low mean value) to the background stellar extinction (a high mean value), and the model is solved with Markov Chain Monte Carlo (MCMC) sampling \citep{2019A&A...624A...6Y}. 


The systematic error in the molecular cloud distance is about 5\%, which could be larger for distant molecular clouds ($>$1 kpc). As discussed in \citet{2019A&A...624A...6Y},  stellar parallax errors ($>$10\%) cloud cause systematic shits (about 10\%) in the distance, but this effect is only significant for far molecular clouds, whose background stars have large relative parallax errors. Other systematic errors arise from the systematic error of Gaia parallaxes, choice of lower CO emission limits for on-cloud stars, and the choice of region used to calculate distances. Unknown systematic errors in the stellar extinction are also responsible for the systematic error of molecular cloud distances. By comparing with the VLBI parallax measurements, \citet{2019A&A...624A...6Y} found a systematic error of $\sim$5\%. \citet{2019ApJ...879..125Z} estimated a similar value ($\sim$5\%) for the systematic error, which were attributed to the photometry errors, stellar models, and systematic Gaia parallax errors. The effect of chemical properties of molecular clouds and dust characteristics on the systematic distance error is unknown.

\section{Results}

\label{sec:result}

\subsection{Molecular cloud samples}

After removing 88 fake molecular clouds due to a bad channel at about $-$24 \kms, DBSCAN identified 18,413 \cofs\ clouds in total. Figure \ref{fig:lbv} displays the distribution of molecular cloud samples in $l$-$V$ and $l$-$b$ spaces. Figure \ref{fig:lbvhist} demonstrates the histogram of molecular cloud samples in the projection of $l$, $b$, and $v$ spaces. 

The local and the Perseus arms are well separated by the radial velocity as shown in panel (a) of Figure \ref{fig:lbvhist}. Molecular cloud samples are roughly homogeneous along the Galactic longitude but show an evident excess above the Galactic plane ($b>0\deg$). This excess is due to  the Cepheus Bubble  \citep{1994ASPC...65...81P,1998ApJ...507..241P} and the warp of the Galactic disk \citep[e.g. ][]{2006PASJ...58..847N,2020ApJS..246....7S}.

 \begin{figure*}[ht!]
 \centering
 \includegraphics[width=0.8  \textwidth]{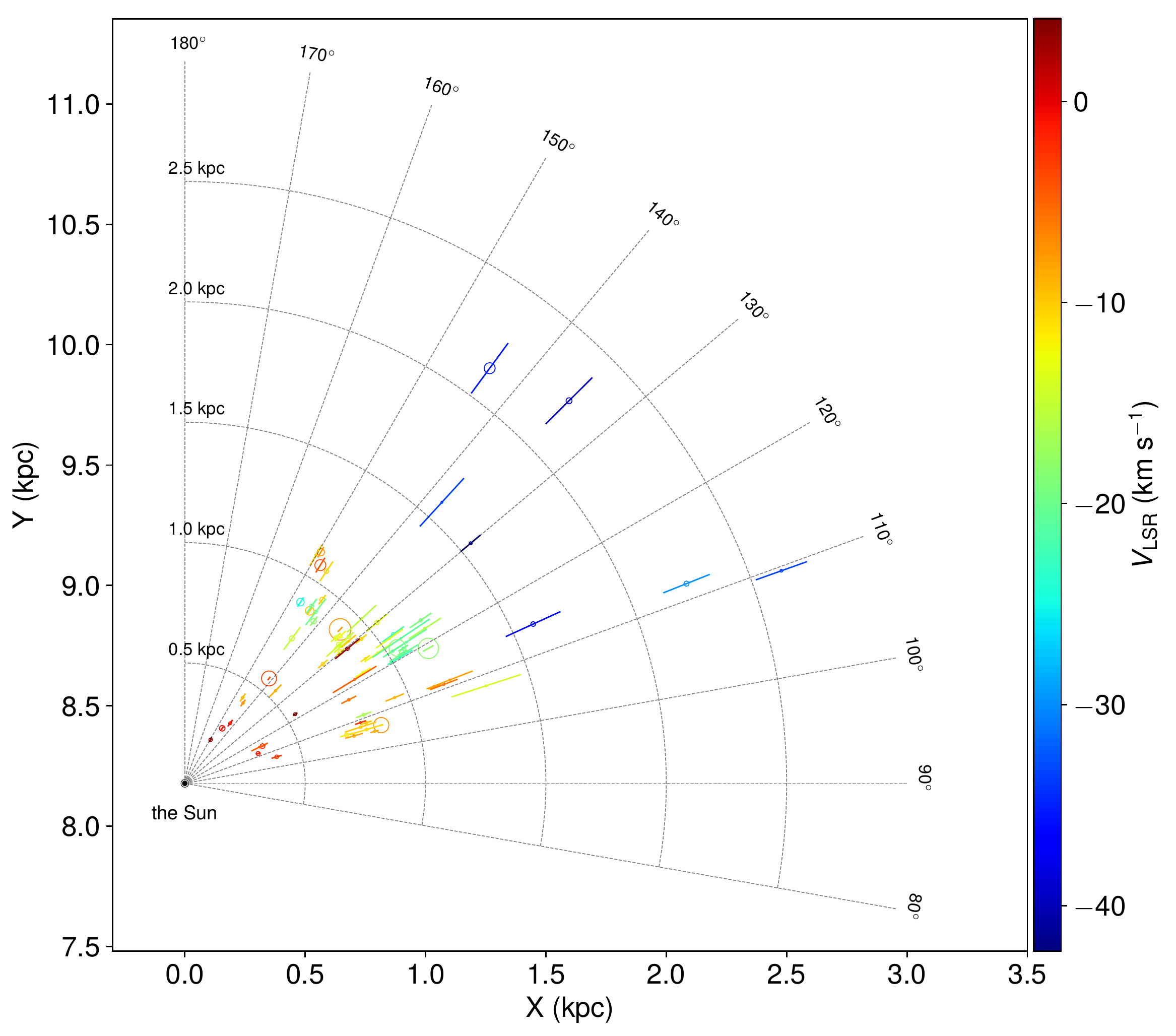} 
\caption{Face on view of 76 molecular clouds with accurate distance measurements in the second Galactic quadrant. The color represents the radial velocity, and the error bar the standard deviation of distance includes the 5\% systematic error. The origin of the coordinate is the Galactic center, and the marker sizes are scaled with the mass.  \label{fig:faceon} }
 \end{figure*}


\subsection{The distance catalog}

In addition to the largest molecular cloud, which is examined below, 76 (of 1677) molecular clouds have distances well determined. The nearest cloud is at 211 pc, while the farthest measured distance is 2631 pc. As examples, we display distances to two molecular clouds G134.7$-$00.3 (\ag) and G112.2$-$01.5 (\av) in Figure \ref{fig:twoclouds}. The angular area of G134.7$-$00.3 is large, and in order to guarantee a small angular separation  between on- and off-cloud stars, we select a fraction of the cloud, and its distance is about 900 pc. G112.2$-$01.5 is relatively small and isolated, and the numbers of on- and off-cloud stars are many fewer than that of G134.7$-$00.3.  Distance figures of all molecular clouds are publicly available on the Harvard Dataverse (\href{https://doi.org/10.7910/DVN/ZX8D8K}{https://doi.org/10.7910/DVN/ZX8D8K}).

Table \ref{Tab:discat} lists distances to 76 molecular clouds. Column (8) is the upper distance cutoff, removing far background stars that is less useful for the distance determination. The total mass of molecular gas (col. 9) is derived with the $^{12}$CO-to-H$_2$ mass conversion factor of X = $2.0\times 10^{20}$ cm$^{-2}$ $\rm(K\  km\ s^{-1})^{-1}$ \citep{2013ARA&A..51..207B} using the \cofs\ flux. Columns (10) lists the maser-parallax-based distances estimated with the model of \citet{2019ApJ...885..131R}.


Figure \ref{fig:faceon} displays the face-on view of the 76 molecular clouds. Evidently, many molecular clouds gather around 1 kpc, and seven molecular clouds  are located in the Perseus arm according to the radial velocity \citep[from $\sim$$-60$  to $\sim$$-30$ \kms, ][]{2020ApJS..246....7S}. G143.7$-$03.3 at $\sim$2.1 kpc, possesses the largest scale height ($\sim$123 pc).





  \begin{figure*}[ht!]
 \centering
 \includegraphics[width=0.95\textwidth]{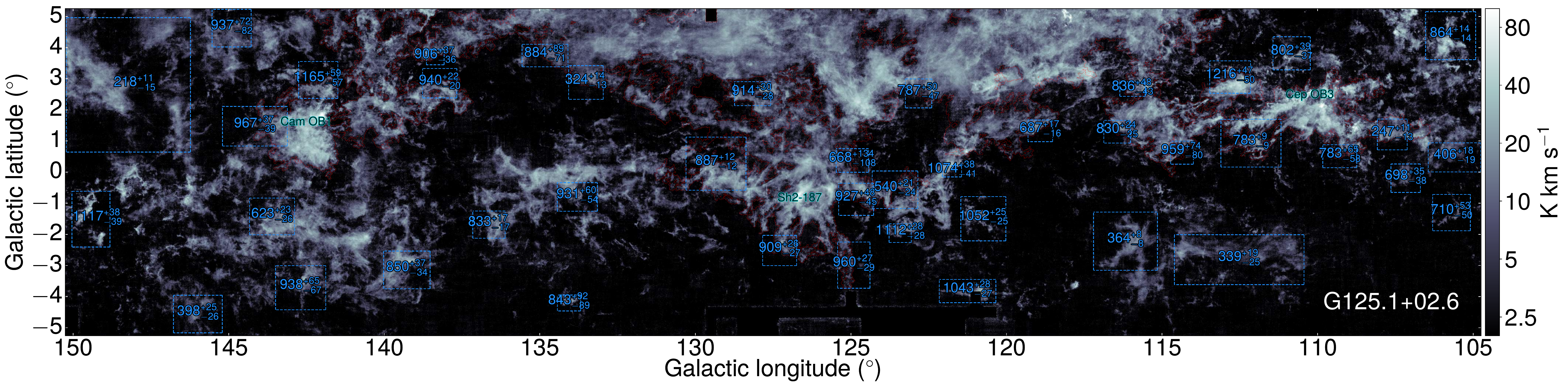} 
\caption{Distances to the largest molecular cloud G125.1+02.6. The edge of the molecular cloud is delineated with the red line. The integrated velocity range is [$-$22.63, 7.55] \kms. The distance is in pc, and distances to many molecular clouds around G125.1+02.6 are also displayed. \label{fig:largest} }
 \end{figure*}

\begin{figure*}[htpb]
\centering
 \includegraphics[width=0.95\textwidth]{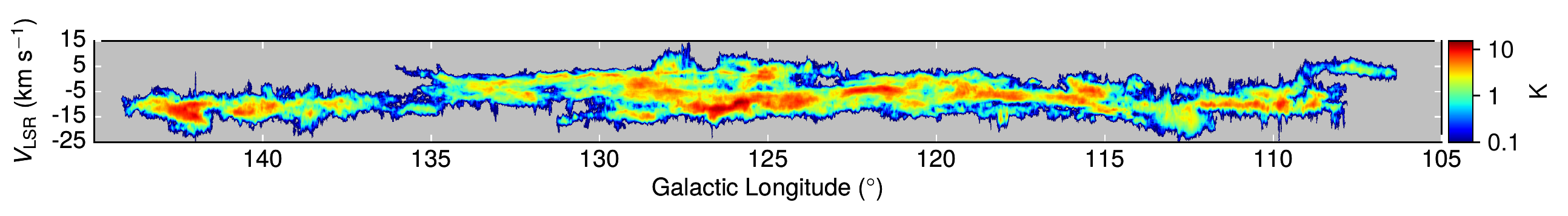} 
\caption{$l$-$V$ diagram of the largest molecular cloud G125.1+02.6. \label{fig:largestlv} }
\end{figure*}

\begin{figure*}[htpb]
\centering
  \includegraphics[width=0.9\textwidth]{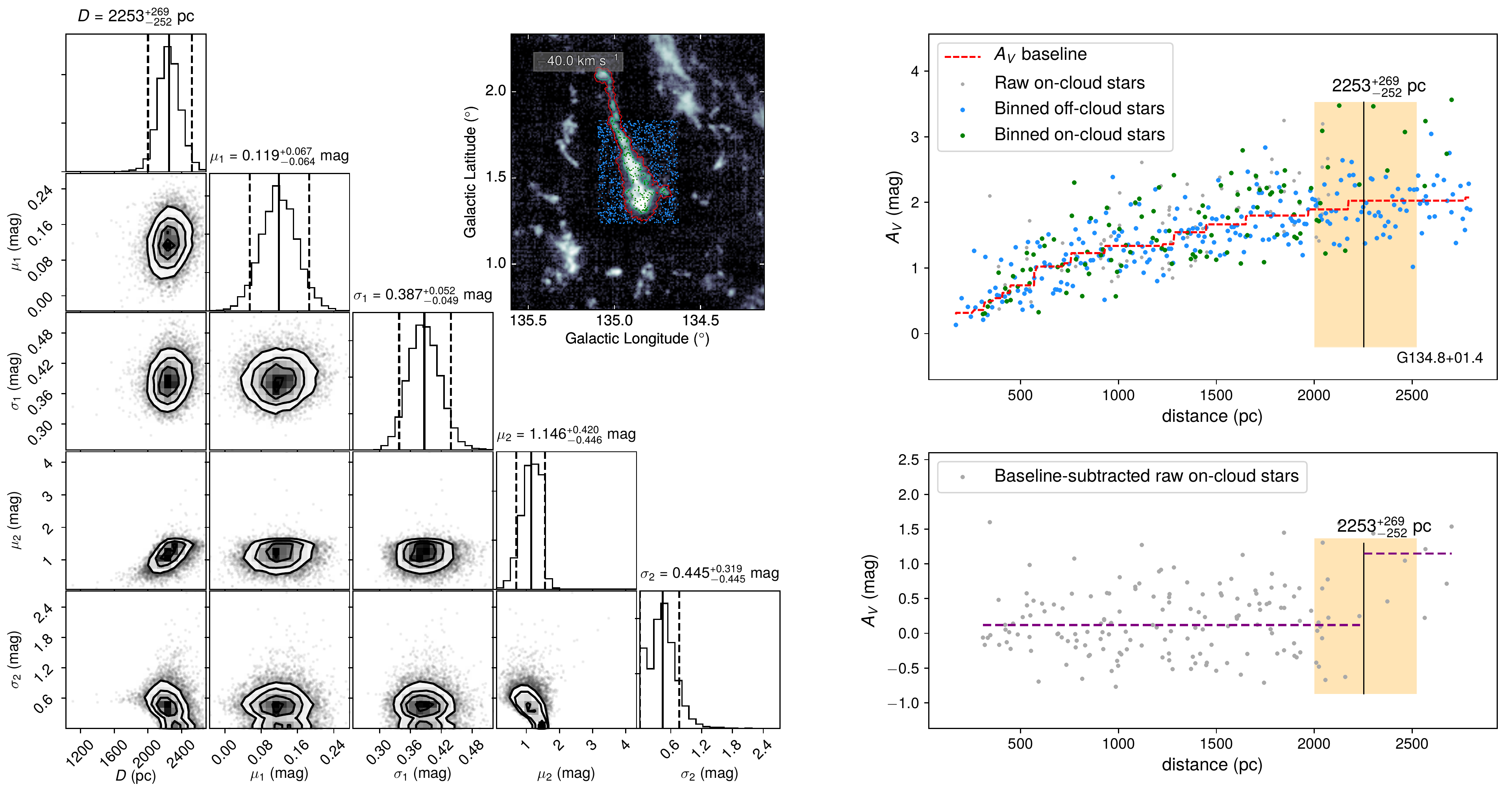}
\caption{Same as Figure \ref{fig:twoclouds} but for molecular cloud G134.8$+$01.4.  G134.8$+$01.4 is part of the W4 \HII\ region. \label{fig:w4} }
\end{figure*}

\subsection{The largest molecular cloud}

The angular area of the largest molecular cloud, G125.1+02.6, is about 107.07 square degrees, which is so large that we have to examine distances to its sub-regions. Most parts of this cloud are above the Galactic plane, occupying a large range in the Galactic longitude, from $\sim$105\deg\ to $\sim$145\deg. The average radial velocity is about $-7.54$ \kms, with a dispersion of 5.03 \kms.  

Distances to G125.1+02.6 are demonstrated in Figure \ref{fig:largest}. The background image is the integrated CO intensity map from $-$22.63 to 7.55 \kms, and see Figure \ref{fig:largestlv} for the $l$-$V$ diagram. Molecular clouds  around G125.1+02.6 show similar distance results. Despite of the contamination of foreground components at $\sim$350 pc, G125.1+02.6 appears to be a coherent structure at about 1 kpc. The length of G125.1+02.6 is about 700 pc, and collectively, the mass of G125.1+02.6 is about $1.5\times10^7$ \msun.



\section{Discussion}
\label{sec:discuss}

\subsection{Comparison with maser-parallax-based distances}

\begin{figure*}[htpb]
\centering
  \includegraphics[width=0.9 \textwidth]{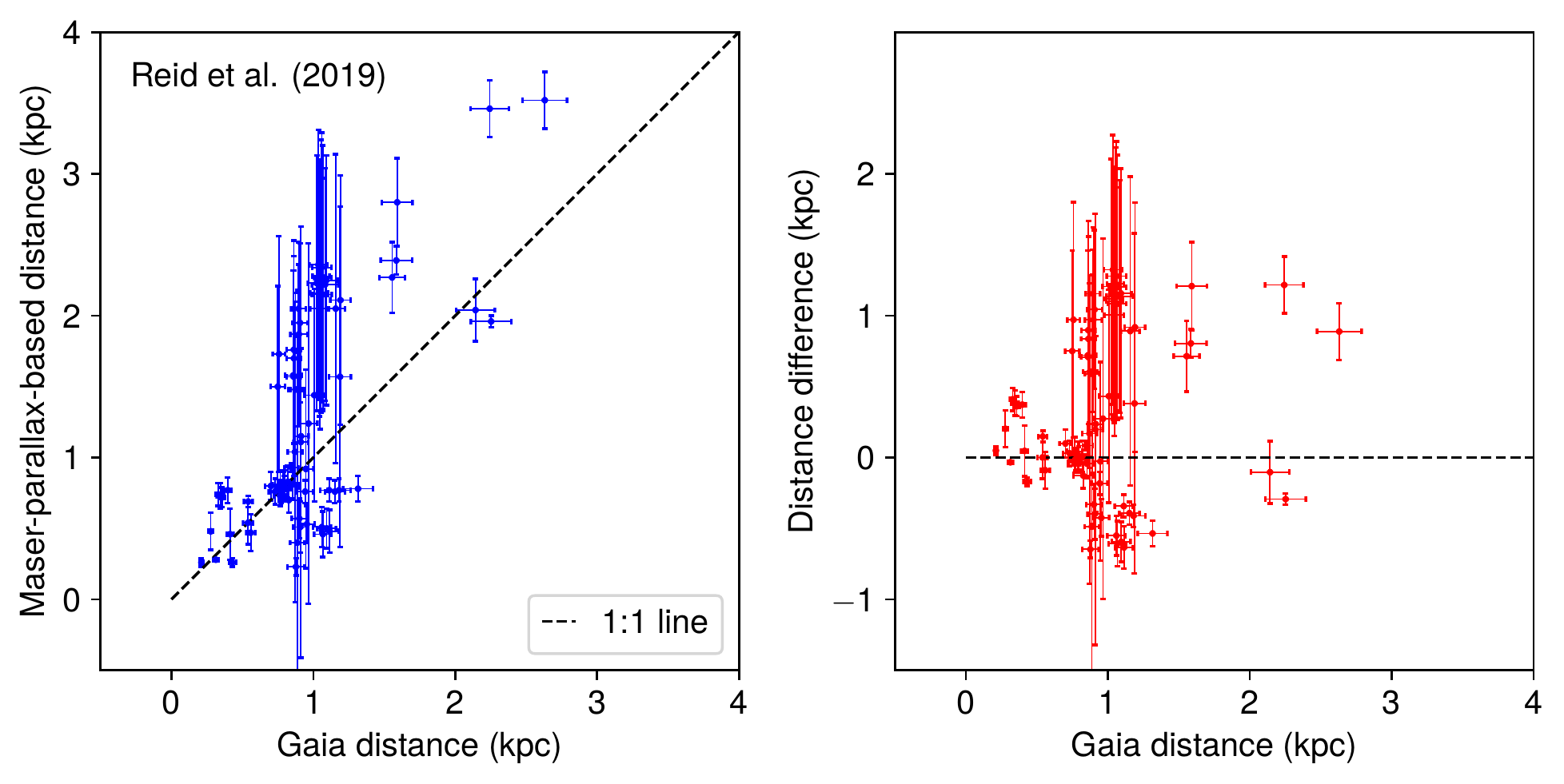}
\caption{Comparison with maser-parallax-based distances derived from the model of \citet{2019ApJ...885..131R}. The error bars represent the standard deviation, and the 5\% systematic error is included for Gaia distances.  \label{fig:kinedis} }
\end{figure*}

In  Figure \ref{fig:kinedis}, we compare our distance results with maser-parallax-based distances derived from \citet{2019ApJ...885..131R}.  


As shown in Figure \ref{fig:kinedis}, for local ($<$1 kpc) molecular clouds, maser-parallax-based distances are generally compatible with the Gaia distances within errors, but for molecular clouds in the Perseus arm, the maser-parallax-based distance is systematically larger by about 600 pc. Even though  \citet{2019ApJ...885..131R} considered the noncircular motions of the Perseus arm \citep[e.g. ][]{1976ApJ...206..114H,2006Sci...311...54X,2020ApJS..246....7S},  maser-parallax-based distances  are still possibly systematically larger for many molecular clouds in the Perseus arm.



\subsection{Comparison with previous distance results}

The largest cloud contains three well-known star-forming regions, Cep OB3,   Sh2-187, and Cam OB1, which are marked in Figure \ref{fig:largest}. The parallax distance of 6.7 GHz methanol masers in Cep OB3 is 830$\pm$20 pc \citep{2016SciA....2E0878X} and the photometric distance of Cep OB3 is 850$\pm$60 pc \citep{1993A&A...273..619M}, both consistent to the Gaia DR2 distance, $\sim$800 pc.  However, the Gaia DR2 distance to the Sh2-187 is about 900 pc, significantly closer than the photometric distance of 1.4 kpc provided by \citep{2007A&A...470..161R}. The Gaia DR2 distance to Cam OB1 is about 1 kpc, consistent with the photometric distance  975$\pm$90 pc \citep{2001AJ....122.2634L}.



The molecular cloud G134.8$+$01.4 is associated with the W4 \HII\ region \citep{1998ApJ...502..265H}. The driven source of W4 is the open cluster IC 1805, whose distance is 2090.6	pc \citep{2018A&A...618A..93C}. As demonstrated in Figure \ref{fig:w4},  the distance to G134.8$+$01.4 is $2253_{-252}^{+269}$ pc, consistent with that of IC 1805 within errors but about 400 pc farther than the distance derived by \citet{2020A&A...633A..51Z} with stellar extinctions and Gaia DR2 parallaxes. In addition to the W4 \HII\ region, distances to two other molecular clouds (see Table \ref{Tab:discat}), L1265 (G115.6$-$02.7) and L1320 (G122.4$-$00.6), are also provided in \citet{2020A&A...633A..51Z}. The distance to L1265 given by \citet{2020A&A...633A..51Z} is about 344 pc, which is about 14 pc smaller than the results of this work. The distance to L1302 of this work is $\sim$1047 pc, which is about 140 pc farther than that given by \citet{2020A&A...633A..51Z}.  Results of W4, L1265 and L1302 indicate that the distances to \citet{2020A&A...633A..51Z} may be systematically smaller ($\sim$10\%) compared with our work. This is possibly due to the usage of off-cloud calibration in the BEEP method.

In addition to L1302, G122.4$-$00.6 also contains a well-known high-mass star-forming region, L1287 ($l=121\fdg30$ and $b=0\fdg66$). The parallax distance of 6.7 GHz methanol maser emission of L1287 is  $929_{-33}^{+ 34}$ pc \citep{2010A&A...511A...2R}, which is about 100 pc closer than the Gaia extinction distance.  Except for the three molecular clouds associated  with L1265, L1302, and W4, we did not find accurate distance references for the rest 73 molecular clouds.

\subsection{Structure of molecular clouds}

 Figure \ref{fig:faceon} shows a clustering of local molecular clouds at about 1 kpc. Taking the largest molecular cloud (Figure \ref{fig:largest}) into account, local molecular clouds form a layer around 1 kpc in the second Galactic quadrant. Evidently, molecular clouds are also present in inter-arm regions.

Due to the limited range in the Galactic longitude, the large-scale structure of molecular clouds is unclear. With the ongoing of the Gaia mission and the MWISP CO survey, distances to   molecular clouds  in the Northern Sky will be measured with higher completeness, which are expected to reveal the large-scale structure of local molecular clouds.

\begin{figure*}[htpb]
\centering
  \includegraphics[width=0.9   \textwidth]{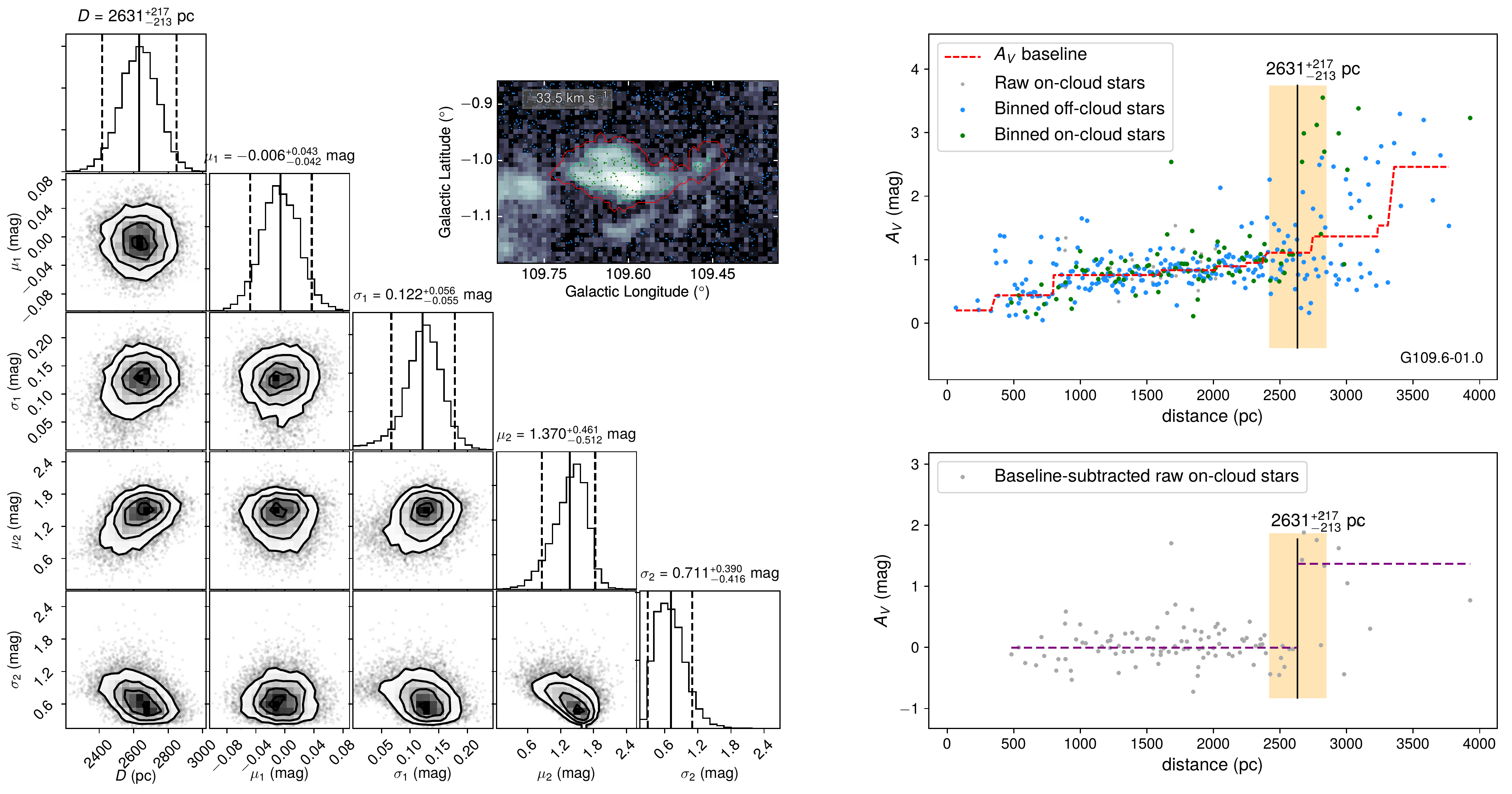}
\caption{ Same as Figure \ref{fig:twoclouds} but for molecular cloud G109.6$-$01.0.  G109.6$-$01.0 is the farthest molecular cloud measured.  \label{fig:farthest} }
\end{figure*}

\subsection{The BEEP method}

In this work, we examined 1677 molecular clouds and measured distances to 76 molecular clouds with the BEEP method. The minimum angular area of molecular clouds that have distance measured is 0.02 square degrees. Molecular clouds with smaller angular areas have too few on-cloud stars to finely trace the extinction variation along lines of sight. In addition to the angular size, however, there are other obstacles responsible for the failure of distance measurements.  

First, distant ($>$2.5 kpc) and optically thick ($>$3.5 mag) molecular clouds are not applicable to the BEEP method. In both cases, there are no sufficient background stars to reveal the extinction jump caused by molecular clouds. See Figure \ref{fig:farthest} for the distance of the farthest molecular cloud, whose background stars are barely enough.


Secondly, nearby ($<$250 pc) molecular clouds usually lack insufficient foreground stars due to the small solid angle, and their distances are usually hard to measure with the BEEP method. Large angular areas are required for those clouds.

Thirdly, molecular clouds with faint CO emission are unable to yield significant extinction jump. For those molecular clouds, large angular areas and relatively simple dust environments  are required. For instance, the BEEP method is still applicable to many diffuse molecular clouds at high Galactic latitudes. 

Occasionally, two molecular clouds occupy approximately the same sky region,  and in this case, the distance usually correspond to the cloud that has higher extinction. 

Despite of these limitations, the BEEP method measures distances to many local molecular clouds  with high accuracy due to its high ability of handling complicated dust environments. With the improvement of  stellar parallax and extinction precision, the BEEP method is expected to   measure distances to many more molecular clouds in the near future.

\section{Summary}
\label{sec:summary}

We have obtained distances to 76 molecular clouds in the second Galactic quadrant, ranging from $\sim$200 to $\sim$2600 pc. 73 of the 76 molecular clouds have their distances accurately determined for the first time. In addition, a local large-scale molecular cloud seems to be a  coherent structure at about 1 kpc, across about 700 pc in the Galactic longitude.


 The maser-parallax-based distances to molecular clouds in the Perseus arm in the second Galactic Quadrant is systematically larger than the Gaia DR2 extinction distances by about 600 pc, possibly due to the streaming motion.

Extension distances to far ($>$2.5 kpc) molecular clouds in the Galactic plane are difficult to measure, due to insufficient background stars and low precision of stellar parallax and extinction data.

 \begin{acknowledgements}
 
We would like to show our gratitude to support members of the MWISP group, Zhiwei Chen, Shaobo Zhang, Min Wang, Jixian Sun, and Dengrong Lu,  and observation assistants at PMO Qinghai station for their long-term observation efforts.  We are also immensely grateful to other member of the MWISP group, Zhibo Jiang, Xuepeng Chen, and Yiping Ao, for their useful discussions. This work was sponsored by the Ministry of Science and Technology (MOST) Grant No. 2017YFA0402701, Key Research Program of Frontier Sciences (CAS) Grant No. QYZDJ-SSW-SLH047, and National Natural Science Foundation of China Grant No. 11773077.

\end{acknowledgements}

\bibliographystyle{aa} 
\bibliography{refGAIADIS} 

 \onecolumn
\onecolumn

\longtab{
\setlength{\tabcolsep}{2.8pt}
\renewcommand{\arraystretch}{1.35}
 \begin{longtable}{cccccccccccccc}
\caption{ Distances to 76 molecular clouds.\label{Tab:discat} } \\
\hline\hline
  Name &  $l$  &   $b$   & $V_{\rm LSR}$  &  Area &  $D_{\rm Gaia}$$^{\rm a}$ &   N$^{\rm b}$     &  $D_{\rm cut}$$^{\rm c}$   &  Mass$^{\rm d}$   &   $D_{\rm Reid2019}$$^{\rm e}$ &$A_V$/$A_G$ & Note \\   
     &  (\deg)  &  (\deg)  & (\kms)  &  deg$^2$ &   (pc)   &     &  (pc) &   ($10^3$ \msun)     &   (kpc)    \\ 
    (1)    &   (2)    &   (3)   &  (4)   &   (5)   &  (6)   &   (7)  &  (8)    &      (9)   &      (10)       &     (11)   &   (12)     \\
\hline
\endfirsthead
\caption{continued.}\\
\hline\hline 
  Name &  $l$  &   $b$   & $V_{\rm LSR}$  &  Area &  $D_{\rm Gaia}$$^{\rm a}$ &   N$^{\rm b}$     &  $D_{\rm cut}$$^{\rm c}$   &  Mass$^{\rm d}$     & $D_{\rm Reid2019}$$^{\rm e}$ & $A_G$/$A_V$ & Note \\   
       &  (\deg)  &  (\deg)  & (\kms)  &  deg$^2$ &   (pc)   &     &  (pc) &   ($10^3$ \msun)     &   (kpc)     \\
    (1)    &   (2)    &   (3)   &  (4)   &   (5)   &  (6)   &   (7)  &  (8)    &      (9)   &      (10)       &     (11)   &   (12)    \\
\hline
\endhead
\hline 
\endfoot
\insertTableNotes
\endlastfoot
 G105.2$+$05.0 &   105.297 &   5.034 &    -8.4  &  0.06 & $ 820_{-  28}^{+  31}$ &    70 & 1500  &   0.5 & $0.83_{-0.11}^{+ 0.11}$ & $A_V$ &         \\  
 G105.6$-$01.3 &   105.686 &  -1.338 &    -8.2  &  0.23 & $ 730_{-  62}^{+  73}$ &   172 & 1500  &   1.6 & $0.76_{-0.09}^{+ 0.09}$ & $A_G$ &         \\  
 G106.1$+$00.5 &   106.100 &   0.590 &    -2.9  &  2.06 & $ 397_{-  33}^{+  26}$ &   360 & 1000  &   3.7 & $0.77_{-0.09}^{+ 0.09}$ & $A_G$ &         \\  
 G106.5$+$01.6 &   106.511 &   1.671 &   -10.8  &  0.27 & $ 787_{- 124}^{+ 142}$ &   119 & 1500  &   1.6 & $0.80_{-0.10}^{+ 0.10}$ & $A_G$ &         \\  
 G106.5$+$04.0 &   106.514 &   4.083 &    -7.0  &  4.89 & $ 853_{-   8}^{+   8}$ &  2639 & 1500  &  75.4 & $0.81_{-0.10}^{+ 0.10}$ & $A_G$ &         \\  
 G106.6$+$01.0 &   106.660 &   1.012 &   -11.7  &  0.52 & $ 701_{-  51}^{+  42}$ &   182 & 1000  &   4.7 & $0.80_{-0.10}^{+ 0.10}$ & $A_G$ &         \\  
 G107.7$+$02.9 &   107.751 &   2.953 &    -9.9  &  0.18 & $ 767_{- 102}^{+  88}$ &   116 & 1500  &   1.8 & $0.81_{-0.10}^{+ 0.10}$ & $A_G$ &         \\  
 G107.9$-$01.3 &   107.949 &  -1.334 &   -14.4  &  0.03 & $1315_{- 288}^{+ 292}$ &    54 & 2500  &   0.3 & $0.78_{-0.09}^{+ 0.09}$ & $A_V$ &         \\  
 G108.4$+$00.3 &   108.483 &   0.362 &    -9.8  &  0.23 & $ 794_{-  39}^{+  42}$ &   139 & 1500  &   1.9 & $0.78_{-0.09}^{+ 0.09}$ & $A_G$ &         \\  
 G109.0$-$00.1 &   109.095 &  -0.156 &    -2.8  &  0.09 & $ 772_{-  40}^{+  40}$ &    89 & 1500  &   0.4 & $0.76_{-0.08}^{+ 0.08}$ & $A_V$ &         \\  
 G109.6$-$01.0 &   109.613 &  -1.027 &   -33.5  &  0.02 & $2631_{- 213}^{+ 217}$ &   106 & 4000  &   1.9 & $3.52_{-0.20}^{+ 0.20}$ & $A_V$ &         \\  
 G110.8$-$01.2 &   110.893 &  -1.233 &    -5.7  &  0.04 & $1152_{- 111}^{+ 118}$ &    80 & 2000  &   0.4 & $0.76_{-0.08}^{+ 0.08}$ & $A_V$ &         \\  
 G110.9$+$03.5 &   110.998 &   3.503 &   -16.6  &  0.19 & $ 797_{-  53}^{+  65}$ &   129 & 1500  &   2.3 & $0.81_{-0.10}^{+ 0.10}$ & $A_G$ &         \\  
 G111.1$-$00.1 &   111.184 &  -0.148 &    -7.0  &  0.14 & $1112_{-  48}^{+  51}$ &   278 & 2000  &   1.6 & $0.77_{-0.08}^{+ 0.08}$ & $A_V$ &         \\  
 G111.3$-$00.5 &   111.324 &  -0.505 &    -8.2  &  0.05 & $1181_{- 166}^{+ 192}$ &    72 & 2500  &   0.4 & $0.77_{-0.08}^{+ 0.08}$ & $A_V$ &         \\  
 G111.7$+$00.0 &   111.706 &   0.008 &   -29.9  &  0.05 & $2243_{- 201}^{+ 197}$ &   114 & 3000  &   8.0 & $3.46_{-0.20}^{+ 0.20}$ & $A_V$ &         \\  
 G112.1$-$02.4 &   112.162 &  -2.435 &    -1.2  &  5.18 & $ 329_{-  13}^{+  12}$ &  1153 & 1000  &   4.9 & $0.74_{-0.08}^{+ 0.08}$ & $A_G$ &         \\  
 G112.2$-$01.5 &   112.249 &  -1.570 &    -8.8  &  0.10 & $ 942_{-  69}^{+  75}$ &    60 & 1300  &   0.6 & $0.76_{-0.08}^{+ 0.08}$ & $A_V$ &         \\  
 G114.5$-$00.1 &   114.573 &  -0.108 &   -36.2  &  0.20 & $1591_{- 206}^{+ 260}$ &   256 & 3000  &   6.6 & $2.80_{-0.31}^{+ 0.31}$ & $A_G$ &         \\  
 G115.6$-$02.7 &   115.611 &  -2.718 &    -2.1  &  4.25 & $ 358_{-  21}^{+  16}$ &   897 &  800  &   6.8 & $0.72_{-0.07}^{+ 0.07}$ & $A_G$ & L1265   \\  
 G115.9$-$00.5 &   115.900 &  -0.559 &    -4.7  &  0.58 & $ 345_{-  56}^{+  67}$ &    81 &  800  &   0.5 & $0.73_{-0.09}^{+ 0.09}$ & $A_G$ &         \\  
 G117.0$+$03.7 &   117.001 &   3.704 &    -5.7  &  0.21 & $ 766_{-  67}^{+  62}$ &    77 & 1000  &   1.4 & $0.73_{-0.07}^{+ 0.07}$ & $A_G$ &         \\  
 G118.9$+$03.0 &   118.998 &   3.033 &   -18.0  &  3.08 & $1158_{-  43}^{+  44}$ &   318 & 2000  & 135.0 & $2.05_{-1.09}^{+ 1.09}$ & $A_G$ &         \\  
 G120.3$-$01.7 &   120.345 &  -1.782 &   -22.7  &  0.02 & $1036_{- 106}^{+ 109}$ &    45 & 2000  &   0.2 & $2.36_{-0.95}^{+ 0.95}$ & $A_V$ &         \\  
 G120.7$-$02.6 &   120.725 &  -2.697 &   -20.2  &  0.03 & $1060_{-  89}^{+ 100}$ &    61 & 2500  &   0.3 & $2.34_{-0.95}^{+ 0.95}$ & $A_V$ &         \\  
 G120.9$-$01.4 &   120.997 &  -1.446 &   -19.5  &  0.49 & $1054_{-  39}^{+  37}$ &   511 & 2000  &   9.3 & $2.28_{-0.96}^{+ 0.96}$ & $A_G$ &         \\  
 G121.4$+$00.5 &   121.460 &   0.532 &    -4.6  &  0.08 & $ 827_{- 187}^{+ 173}$ &    30 & 1500  &   0.3 & $0.70_{-0.09}^{+ 0.09}$ & $A_G$ &         \\  
 G121.7$-$02.2 &   121.768 &  -2.279 &   -19.9  &  0.12 & $1068_{-  41}^{+  41}$ &   149 & 2000  &   1.1 & $2.27_{-0.93}^{+ 0.93}$ & $A_V$ &         \\  
 G121.8$-$00.4 &   121.887 &  -0.452 &   -12.5  &  0.16 & $ 871_{-  66}^{+  68}$ &    71 & 1500  &   1.1 & $1.04_{-1.06}^{+ 1.06}$ & $A_G$ &         \\  
 G122.0$-$00.5 &   122.087 &  -0.550 &     4.1  &  0.48 & $ 541_{-  15}^{+  15}$ &   187 & 1000  &   2.2 & $0.69_{-0.04}^{+ 0.04}$ & $A_G$ &         \\  
 G122.4$-$02.3 &   122.470 &  -2.372 &   -20.6  &  0.10 & $1024_{-  94}^{+  75}$ &   185 & 2000  &   1.0 & $2.23_{-0.90}^{+ 0.90}$ & $A_V$ &         \\  
 G122.4$-$00.6 &   122.500 &  -0.695 &   -17.9  &  3.77 & $1047_{-  15}^{+  15}$ &  3206 & 1500  &  91.1 & $2.15_{-0.95}^{+ 0.95}$ & $A_G$ & L1287,L1302 \\  
 G122.5$-$02.8 &   122.519 &  -2.866 &   -20.5  &  0.02 & $1092_{- 175}^{+ 181}$ &    33 & 2000  &   0.3 & $2.25_{-0.88}^{+ 0.88}$ & $A_V$ &         \\  
 G122.8$-$02.3 &   122.843 &  -2.303 &   -16.3  &  0.12 & $1188_{- 138}^{+ 147}$ &    51 & 2000  &   1.1 & $1.57_{-1.20}^{+ 1.20}$ & $A_G$ &         \\  
 G123.8$-$03.1 &   123.880 &  -3.128 &   -20.9  &  0.02 & $1083_{- 234}^{+ 241}$ &    30 & 2500  &   0.2 & $2.22_{-0.82}^{+ 0.82}$ & $A_V$ &         \\  
 G124.0$+$00.5 &   124.081 &   0.579 &   -13.0  &  0.07 & $ 966_{- 110}^{+ 122}$ &    68 & 2000  &   0.5 & $1.24_{-1.27}^{+ 1.27}$ & $A_V$ &         \\  
 G124.6$-$01.4 &   124.652 &  -1.452 &   -19.2  &  0.17 & $1191_{- 105}^{+  97}$ &   115 & 2500  &   2.8 & $2.11_{-0.88}^{+ 0.88}$ & $A_G$ &         \\  
 G124.6$-$01.9 &   124.659 &  -1.986 &   -11.1  &  0.08 & $ 912_{-  38}^{+  39}$ &   144 & 2000  &   0.6 & $1.11_{-1.52}^{+ 1.52}$ & $A_V$ &         \\  
 G125.2$-$04.0 &   125.297 &  -4.001 &   -13.3  &  0.03 & $1044_{- 121}^{+ 130}$ &    75 & 2500  &   0.3 & $2.05_{-0.76}^{+ 0.76}$ & $A_V$ &         \\  
 G125.5$-$02.4 &   125.597 &  -2.410 &   -22.5  &  0.30 & $1065_{-  83}^{+ 108}$ &   204 & 2500  &   3.3 & $2.15_{-0.82}^{+ 0.82}$ & $A_G$ &         \\  
 G129.3$-$01.6 &   129.311 &  -1.646 &   -10.6  &  0.26 & $ 947_{-  52}^{+  49}$ &   321 & 1500  &   2.6 & $0.92_{-0.70}^{+ 0.70}$ & $A_V$ &         \\  
 G129.6$-$00.1 &   129.631 &  -0.145 &     3.1  &  0.42 & $ 877_{- 100}^{+  95}$ &   184 & 1500  &   3.5 & $0.23_{-0.06}^{+ 0.06}$ & $A_G$ &         \\  
 G129.9$-$02.1 &   129.945 &  -2.148 &   -13.5  &  0.60 & $1041_{- 100}^{+  99}$ &   308 & 2000  &   6.0 & $2.16_{-0.52}^{+ 0.52}$ & $A_G$ &         \\  
 G130.0$-$04.4 &   130.026 &  -4.439 &   -42.3  &  0.08 & $1555_{-  97}^{+ 104}$ &   163 & 2500  &   2.2 & $2.27_{-0.25}^{+ 0.25}$ & $A_V$ &         \\  
 G130.1$+$00.5 &   130.143 &   0.594 &   -11.5  &  0.68 & $ 845_{-  40}^{+  41}$ &   306 & 1500  &   7.9 & $0.93_{-0.03}^{+ 0.03}$ & $A_G$ &         \\  
 G130.8$-$01.0 &   130.866 &  -1.000 &   -10.3  &  0.51 & $ 758_{-  46}^{+  44}$ &   156 & 1500  &   2.5 & $1.73_{-0.83}^{+ 0.83}$ & $A_G$ &         \\  
 G130.9$+$01.2 &   130.970 &   1.208 &   -10.0  &  0.06 & $ 901_{-  69}^{+  77}$ &    49 & 1500  &   0.4 & $0.57_{-0.14}^{+ 0.14}$ & $A_V$ &         \\  
 G131.2$-$00.5 &   131.215 &  -0.521 &   -14.3  &  0.07 & $ 863_{-  66}^{+  73}$ &    97 & 1500  &   0.4 & $1.70_{-0.83}^{+ 0.83}$ & $A_V$ &         \\  
 G131.8$+$00.0 &   131.838 &   0.093 &   -15.6  &  0.45 & $ 914_{-  46}^{+  48}$ &   254 & 1500  &   3.8 & $1.15_{-0.62}^{+ 0.62}$ & $A_V$ &         \\  
 G131.8$-$02.5 &   131.841 &  -2.556 &   -14.5  &  0.10 & $ 859_{- 127}^{+ 122}$ &   138 & 2000  &   0.5 & $1.58_{-0.74}^{+ 0.74}$ & $A_V$ &         \\  
 G132.0$-$01.1 &   132.035 &  -1.123 &   -14.0  &  0.84 & $ 861_{-  44}^{+  45}$ &   590 & 1500  &   6.8 & $1.57_{-0.75}^{+ 0.75}$ & $A_V$ &         \\  
 G132.9$-$04.5 &   132.972 &  -4.536 &   -15.5  &  0.05 & $1007_{- 148}^{+ 137}$ &    45 & 2000  &   0.3 & $1.44_{-0.75}^{+ 0.75}$ & $A_V$ &         \\  
 G133.4$-$02.0 &   133.440 &  -2.068 &   -12.6  &  0.63 & $ 890_{-  54}^{+  57}$ &   341 & 1500  &   4.5 & $1.48_{-0.70}^{+ 0.70}$ & $A_V$ &         \\  
 G134.0$-$04.0 &   134.067 &  -4.075 &   -13.8  &  0.26 & $ 874_{-  48}^{+  45}$ &   354 & 1500  &   2.8 & $1.48_{-0.68}^{+ 0.68}$ & $A_V$ &         \\  
 G134.7$-$00.3 &   134.761 &  -0.382 &    -7.6  & 11.02 & $ 908_{-  21}^{+  22}$ &  1075 & 1500  & 151.7 & $0.51_{-0.18}^{+ 0.18}$ & $A_G$ &         \\  
 G134.8$+$01.4 &   134.878 &   1.493 &   -40.0  &  0.09 & $2253_{- 252}^{+ 269}$ &   163 & 2800  &  10.9 & $1.96_{-0.04}^{+ 0.04}$ & $A_V$ & W4      \\  
 G135.6$+$03.5 &   135.630 &   3.527 &    -8.6  &  0.28 & $ 539_{-  51}^{+  63}$ &   156 & 1000  &   0.8 & $0.54_{-0.15}^{+ 0.15}$ & $A_V$ &         \\  
 G137.5$-$03.3 &   137.553 &  -3.390 &   -33.6  &  0.02 & $1585_{- 230}^{+ 270}$ &    36 & 2500  &   0.5 & $2.39_{-0.10}^{+ 0.10}$ & $A_V$ &         \\  
 G141.2$-$02.0 &   141.256 &  -2.038 &    -3.5  & 16.87 & $ 559_{-  12}^{+  12}$ &  1747 & 1000  &  72.4 & $0.47_{-0.13}^{+ 0.13}$ & $A_G$ &         \\  
 G141.5$-$03.4 &   141.520 &  -3.424 &   -18.1  &  1.18 & $ 862_{-  38}^{+  43}$ &   365 & 1500  &   9.4 & $1.76_{-0.66}^{+ 0.66}$ & $A_G$ &         \\  
 G142.6$-$03.6 &   142.658 &  -3.627 &   -19.7  &  0.45 & $ 898_{- 125}^{+ 119}$ &   133 & 1500  &   3.0 & $1.87_{-0.65}^{+ 0.65}$ & $A_G$ &         \\  
 G143.0$+$00.7 &   143.021 &   0.784 &    -0.5  &  2.73 & $ 313_{-  27}^{+  30}$ &   129 &  500  &   2.1 & $0.28_{-0.01}^{+ 0.01}$ & $A_V$ &         \\  
 G143.2$+$03.0 &   143.202 &   3.009 &   -10.4  &  0.68 & $ 954_{-  40}^{+  41}$ &   404 & 2000  &   8.1 & $0.53_{-0.13}^{+ 0.13}$ & $A_G$ &         \\  
 G143.5$-$01.5 &   143.542 &  -1.598 &   -14.9  &  1.24 & $ 749_{- 102}^{+ 105}$ &   379 & 1500  &   8.5 & $1.50_{-0.71}^{+ 0.71}$ & $A_G$ &         \\  
 G143.7$-$03.3 &   143.707 &  -3.342 &   -35.3  &  0.45 & $2143_{- 249}^{+ 243}$ &    74 & 3000  &  38.6 & $2.04_{-0.22}^{+ 0.22}$ & $A_V$ &         \\  
 G144.0$-$03.8 &   144.078 &  -3.824 &    -7.4  &  0.95 & $ 413_{-  30}^{+  26}$ &   154 & 1000  &   1.2 & $0.46_{-0.18}^{+ 0.18}$ & $A_G$ &         \\  
 G144.0$+$04.2 &   144.096 &   4.220 &    -9.9  &  2.46 & $ 887_{-  26}^{+  29}$ &   928 & 1500  &  23.1 & $0.40_{-1.45}^{+ 1.45}$ & $A_G$ &         \\  
 G144.4$-$02.5 &   144.438 &  -2.595 &   -20.0  &  0.61 & $ 906_{-  66}^{+  65}$ &    99 & 1100  &   4.3 & $1.95_{-0.56}^{+ 0.56}$ & $A_G$ &         \\  
 G145.8$+$03.4 &   145.817 &   3.428 &    -0.4  & 11.64 & $ 277_{-  14}^{+  14}$ &   713 &  500  &  10.1 & $0.48_{-0.13}^{+ 0.13}$ & $A_G$ &         \\  
 G146.0$-$04.2 &   146.013 &  -4.267 &    -9.2  &  1.00 & $ 431_{-  29}^{+  37}$ &   235 &  800  &   1.4 & $0.26_{-0.03}^{+ 0.03}$ & $A_V$ &         \\  
 G146.2$-$02.3 &   146.273 &  -2.388 &   -10.5  &  0.52 & $1060_{-  91}^{+  89}$ &   172 & 2000  &   5.9 & $0.51_{-0.14}^{+ 0.14}$ & $A_G$ &         \\  
 G147.4$-$04.0 &   147.461 &  -4.017 &   -24.9  &  1.94 & $ 895_{-  39}^{+  40}$ &   540 & 1500  &  17.9 & $2.05_{-0.31}^{+ 0.31}$ & $A_G$ &         \\  
 G148.1$-$00.2 &   148.160 &  -0.237 &    -4.2  &  1.97 & $1067_{-  66}^{+  63}$ &   689 & 1500  &  38.9 & $0.46_{-0.16}^{+ 0.16}$ & $A_G$ &         \\  
 G149.4$+$03.1 &   149.408 &   3.188 &     3.1  &  5.28 & $ 211_{-  11}^{+  13}$ &   281 &  500  &   3.4 & $0.26_{-0.03}^{+ 0.03}$ & $A_G$ &         \\  
 G149.5$-$01.0 &   149.535 &  -1.088 &    -7.7  &  0.50 & $1114_{-  44}^{+  44}$ &   364 & 2000  &  16.9 & $0.48_{-0.15}^{+ 0.15}$ & $A_G$ &         \\  
 G150.1$-$01.4 &   150.131 &  -1.417 &   -10.3  &  0.05 & $1094_{-  93}^{+  93}$ &    98 & 2000  &   0.6 & $0.50_{-0.14}^{+ 0.14}$ & $A_V$ &         \\  
 \hline
\end{longtable}
  }
 
\begin{TableNotes}
\footnotesize
\item [a]  The error of distances is the 95\% HPD, about two standard derivations for Gaussian distributions, and the 5\% systematic error is not included. The machine-readable table is publicly accessible on the Harvard Dataverse (\href{https://doi.org/10.7910/DVN/MNKOCH}{https://doi.org/10.7910/DVN/MNKOCH}).   
\item [b]  Total number of on-cloud stars. 
\item [b]   Background stars farther than $D_{\rm cut}$ are excluded in the distance measurement.
\item [d]   Total mass of molecular gas in molecular clouds estimated with the $^{12}$CO-to-H$_2$ mass conversion factor of X = $2.0\times 10^{20}$ cm$^{-2}$ $\rm(K\  km\ s^{-1})^{-1}$ \citep{2013ARA&A..51..207B}, which only takes account of CO-bright components. 
\item [e] Maser-parallax-based distances derived from the model of \citet{2019ApJ...885..131R}. The error is the standard derivation.
\end{TableNotes}

\end{document}